%% file: mots.tex
\documentclass[a4paper,twoside]{article}

\usepackage[nolist]{acronym}
\usepackage{multicol}

\usepackage{amsfonts}

\usepackage{booktabs}

\usepackage{enumitem}% http://ctan.org/pkg/enumitem

\usepackage{wasysym}

\usepackage{listings}
\usepackage[export]{adjustbox}

\usepackage{tikz}
\usepackage{epstopdf}
\usetikzlibrary{positioning}
\usetikzlibrary{fit}
\usetikzlibrary{graphs}
\usetikzlibrary{backgrounds}
\usetikzlibrary{quotes}
\usetikzlibrary{shapes}
\usetikzlibrary{calc}
\usetikzlibrary{arrows}
\graphicspath{ {./figures/} }

\usepackage{epsfig}
\usepackage{subfigure}
\usepackage{calc}
\usepackage{amssymb}
\usepackage{amstext}
\usepackage{amsthm}
\usepackage{pslatex}
\usepackage{apalike}
\usepackage{SCITEPRESS}
\begin{document}
\begin{acronym}[NNNNNNN]
	\input{00acronyms.tex}
\end{acronym}

% =========================================================================== %
\title{Towards Understanding Man-on-the-Side Attacks (MotS) in SCADA Networks}

\author{\authorname{Peter Maynard and Kieran McLaughlin}
\affiliation{Centre for Secure Information Technology\\ Queen's University Belfast, UK}
\email{\{p.maynard,kieran.mclaughlin\}@qub.ac.uk}}

% =========================================================================== %

% =========================================================================== %

\abstract{We describe a new class of packet injection attacks called \ac{MotS}, previously only seen where state actors have ``compromised'' a number of telecommunication companies. \ac{MotS} injection attacks have not been widely investigated in scientific literature, despite having been discussed by news outlets and security blogs. \ac{MotS} came to attention after the Edward Snowden revelations, which described large scale pervasive monitoring of the Internet's infrastructure. For an advanced adversary attempting to interfere with IT connected systems, the next logical step is to adapt this class of attack to a smaller scale, such as enterprise or critical infrastructure networks. \ac{MotS} is a weaker form of attack compared to a \ac{MitM}. A \ac{MotS} attack allows an adversary to read and inject packets, but not modify packets sent by other hosts. This paper presents practical experiments where we have implemented and performed \ac{MotS} attacks against two testbeds: 1) on HTTP connections, by redirecting a victim to a host controlled by an adversary; and 2) on an Industrial Control network, where we inject falsified command responses to the victim. In both cases, the victims accept the injected packets without generating a suspiciously large number of unusual packets on the network. We then perform an analysis of three leading Network \acp{IDS} to determine whether the attacks are detected, and discuss mitigation methods.}
% =========================================================================== %
\acresetall 

\onecolumn \maketitle \normalsize \vfill

% =========================================================================== %
\section{\uppercase{Introduction}}
% =========================================================================== %

\let\thefootnote\relax\footnotetext{Reproducible experiments can be found at: \\\indent\url{https://github.com/PMaynard/mots}}

% Snowden - mots + others disclosed, www/companies reacted with TLS; ICS Can't 
The Snowden revelations have been a driving factor towards a more secure digital world. One clear example of this is that TLS has become widely deployed on systems which previously did not use encryption. This paper builds upon leaked information regarding \ac{MotS} attacks. \ac{MotS} has traditionally been considered only achievable by a global pervasive adversary \cite{trammell_confidentiality_2015}, such as a state actor, who is capable of using their global timing advantage to inject forged packets before a victim receives legitimate packets, which are then correctly discarded by the target. In contrast, the experiments presented here investigate a localised implementation of \ac{MotS} attacks against two protocols, HTTP and \ac{IEC104}. \ac{IEC104} is widely used within Europe and Asia for controlling and monitoring critical infrastructure e.g. gas, electricity, etc. With state actors known to be targeting infrastructure such as electricity grids, our motivation is to better understand how such adversaries may adopt \ac{MotS} attack techniques, which they are known to have used in the past in different circumstances.

% NIST propose improved security for ICS just after Snowden.
In 2013, a NIST report \cite{forrester_developing_2013} advocated the improvement of industrial networks through the use of zero-trust networks. DeCusatis et.el \cite{decusatis_implementing_2016} discussed how to implement a zero-trust network, which would mitigate this class of attacks. DeCusatis implemented a policy engine and gateway which prevents unauthorised packets from being accepted by a victim, as well as ensuring that the attacker does not receive identifiable information when their probe is dropped. However, within in the \ac{ICS} domain, it is not possible to quickly deploy radical changes as seen with TLS. Information from the National Grid \cite{national_grid_response_2013} shows that this major upgrade of critical networks is not possible in the time frame laid out.
% Why ICS can't react, does it need to?
Due to the long life cycle of \ac{ICS} equipment and the criticality of the physical systems they control, many security recommendations for IT networks are not suitable for \ac{OT} networks. As a consequence \ac{OT} networks frequently rely on unencrypted protocols. Threats are typically mitigated with the use of \acp{VPN}, air gapped networks, network segmentation, firewalls, and \acp{DMZ}. 

However, industrial control networks are often targeted by state actors, as was the case with the TRISIS malware \cite{johnson_attackers_2017}. This was a \ac{RAT} that targeted Triconex \ac{SIS} devices, and triggered a shut down of an industrial site in the Middle East. Another example is CRASHOVERRIDE \cite{dragos_crashoverride_2017}, which performed network enumeration and exploitation of \ac{SCADA} protocols including \ac{IEC104}. This malware has been linked to the Ukrainian power outages in 2016. Therefore, it is logical to assume that \ac{MotS} has, or is likely to be, deployed by similar adversaries into an industrial network at some point in time.

% Contributions

Many papers that propose novel network \acp{IDS}, tend verify their effectiveness using the outdated KDD'99 dataset or common attacks (i.e. \ac{MitM}, replay and injection \cite{maynard_towards_2014,green_significance_2017}). We provide a modern technique which should be considered when validating intrusion detection methods. We investigate the detection responsiveness of the current state of the art network \acp{IDS} (Snort, Zeek, Suricata) when confronted with \ac{MotS}, for both HTTP and \ac{IEC104}. We provide reproducible experiments and a packet captures of our attacks, so they may be used to validate other \acp{IDS}. 

The paper continues with an overview of related work (\S\ref{sec:related}) and background (\S\ref{background}), which details what required for a \ac{MotS} attack, and introduces ICS/SCADA networking architecture. Following on to the experimentation methodology (\S\ref{sec:method}), and experimental results (\S\ref{sec:exp}). Finally, a review of current detection systems (\S\ref{sec:detection}) and conclusion (\S\ref{sec:conclude}). 

% =========================================================================== %
\section{\uppercase{Related Work}}
\label{sec:related}
% =========================================================================== %

% Off the path and quick history on TCP inject
In 2012, Gilad et al. \cite{gilad_off-path_2012} proved that it is possible to inject packets into a TCP stream, by using browser based malware to identify the IP-ID, then spoof a request. Despite reports from IETF RFC 6528, which state that TCP segment injection is mitigated by having the majority of data needed to perform this ``not well known'' to the attacker. While the IETF suggests VPN or TCP Authentication Option (2010) to prevent injection. At the same, time the IETF acknowledges the TCP reset vulnerability is a known issue. As of 2019, Alexander et al. \cite{alexander_detecting_2019} can identify the IP-port four-tuple representing an active TCP connection, by taking advantage of side-channels within the Linux kernel 4.0 and above. Packet injection and TCP resets are a widely known issue, with current mitigations easily bypassed. As indicated in the introduction, the existing state of the art IDSs for ICS networks are not validated against these types of attacks. 

The Chinese Great Firewall exploits the TCP reset vulnerability, and is discussed by Weaver et al. \cite{weaver_detecting_2009} (2009) who propose mitigations. Also in 2012, Victor Julien added a detection rule for overlapping data to the open source network \ac{IDS} Suricata. Essentially, detecting TCP injection and \ac{MotS} attacks as seen in the detection section. In 2014, Gilad et al. \cite{gilad_off-path_2014} improved upon their previous TCP injection attacks, so it does not require the use of web malware. They also detail the history of TCP injection attacks. While Gilad et al. focus on injecting attacks over the internet, we choose to focus on exploiting a target on the local network, assuming that it has been compromised by a skilled adversary. 

% Snowden happens
Between 2013 and 2014, The Guardian and The Intercept \cite{schneier_attacking_2013,gallagher_how_2014} discuss a nation state operation called ``QUANTUM Inject'', which uses their wide scope to monitor and inject packets into TCP streams. 
% Uptake of mitigations and explinations.
Following this media attention, the security company Fox-IT performed a deep dive into QUANTUM Inject \cite{haagsma_deep_2015}. They developed a PoC and released patches for a number of open source network \acp{IDS}.
% IETF get in on the action.
In 2015, a report from the IETF \cite{trammell_confidentiality_2015} discussed a threat model of a passive pervasive attacker, explicitly stating \ac{MotS} as one example.
% hjelmvik talks.
The company NetReSec, which performs network security training and develops a packet analyser tool, have written a number of times about \ac{MotS} between 2015-16 \cite{hjelmvik_covert_2015}. They attempted to raise awareness of the attack, and developed additional detection tools as well as identifying TCP injection attacks from China.

% Exactly MotS stuff 
Marczak et al. \cite{marczak_analysis_2015} likened the Distributed \ac{DOS} attack from ``China's Great Cannon'' to the QUANTUM system. In that the techniques used to inject packets are similar to the QUANTUM Inject method. Finally, the closest related work the authors identified was by Nakibly et al. \cite{nakibly_website-targeted_2016}, who detail \ac{MotS} attacks being used to inject Javascript, HTTP redirections and malware dropping. They performed a study to analyse incoming and outgoing connections of four university institutions, and identified several different threat groups. They proposed a few \ac{MotS} detection methods, which use the IP features \ac{TTL} and IP ID, and timing analysis to identify \ac{MotS} in the wild. The authors are unaware of any papers detailing \ac{MotS} class of attack on industrial networks, and are certain there is no academic literature that provides reproducible experiments and packet captures that may be used to validate new IDSs.   

% =========================================================================== %
\section{\uppercase{Background}}
\label{background}
% =========================================================================== %

% --------------------------------------------------------------------------- %
% \subsection{\acf{MotS}}
% --------------------------------------------------------------------------- %
% How is this not a MitM? Aka what are the advantages or disadvantages of it compared to MitM?
\ac{MotS}, like SYN-flooding and DNS-Spoofing attacks, are known as \textit{off-path}. This differs from traditional \ac{MitM} attacks \cite{conti_survey_2016} which are considered \textit{on-path}. An \textit{on-path} attacker needs to control the links between the victim and host, while an \textit{off-path} attacker does not. Although \ac{MotS} is a weaker class of attack than \ac{MitM}, it is capable of packet injection, without having to maintain connections as a proxy. Previous work by the authors  \cite{maynard_towards_2014} which used \ac{ARP} spoofing to maintain an \textit{on-the-path} attack, required a large number of \ac{ARP} packets, as well as the need to continuously forward packets to the victim. Depending on the network and host, this may be easily detectable. It can also be resource intensive. On the other hand, a \ac{MotS} attack can be performed with a minimum of one forged packet. Also, \ac{MotS} can not be mitigated by locking down switch ports using \ac{MAC} \acp{ACL}, which can mitigate \ac{MitM}.   

% characterise mots attack
A \ac{MotS} attack can be successfully performed if the attacker can: a) observe a victim's request; b) adequately replicate a response that the victim would accept; and c) transmit a response to the victim quicker than legitimate response. If these three requirements are met, then an adversay can perform this attack. Fundamentally, this attack exploits the design of TCP (RFC 793), whereby it is normal for TCP segments to arrive more than once, and that the first segment that arrives is accepted. The attacker needs to be in a position to know the four-tuple (Local IP address, Local Port, Remote IP address, Remote Port) that identifies the target connection, along with the next sequence number expected by the target.

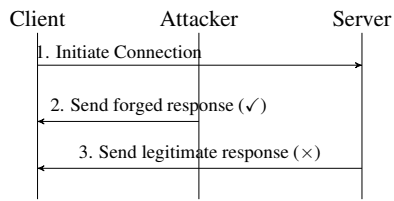
\begin{figure}
  \centering
  \resizebox{.7\linewidth}{!}{\input{figures/mots-sequence}}
  \caption{Sequence Diagram of \acf{MotS} attack.}
  \label{fig:mots-sequence}
\end{figure}

Figure~\ref{fig:mots-sequence} shows the sequence diagram of a \ac{MotS} attack, consisting of a Client, Server and Attacker. First, the Client initiates a connection with the server and a TCP handshake is performed, followed by a request to the server. The attacker is positioned within the network so that they can view this request (the attacker does not need to view the handshake). The attacker creates and sends a forged response, which spoofs the four-tuple information, including the TCP sequence number. The forged packet is accepted by the client. Subsequently the legitimate response arrives and is dropped by the client, as it has already acknowledged the forged packet.  

% --------------------------------------------------------------------------- %
\subsection{Industrial Control System Networks}
\label{subsec:icsnet}
% --------------------------------------------------------------------------- %

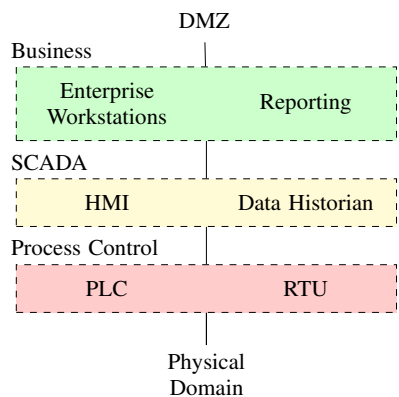
\begin{figure}
	\centering
	\resizebox{.7\linewidth}{!}{\input{figures/highlevel-ics-overview}}
	\caption{A high level network diagram of an \acf{ICS}.}
	\label{fig:scada-network}
\end{figure}

The terms \acf{ICS} and \acf{SCADA} are often used interchangeably. However, \ac{SCADA} may be considered part of an \ac{ICS}, which is used to monitor and control physical devices. Figure~\ref{fig:scada-network} shows a high level network diagram of an \ac{ICS} network, that highlights the different network segments, which have limited access between each other.

In the diagram, the \ac{SCADA} enclave is located in the middle and provides information to the business network, as well as monitoring the process control enclave. A \ac{HMI} may be used to monitor the physical process, while a Data Historian maintains a record of plant operations. The process control enclave contains a number of devices used to govern physical instruments and actuators that interact with the physical domain. Typical devices include \acp{PLC} and \acp{RTU} which are rugged embedded devices designed for a distinct task. The network may consist of serial links, Ethernet, and IP, as well as field bus protocols; one such protocol is \ac{IEC104} \cite{international_electrotechnical_commission_en_2006}. If \ac{MotS} were to be performed on ICS network, the risks to the system may be: 

\begin{itemize}[noitemsep]
     \item \textbf{Unauthorised access to information}: 
      Redirect information to the adversary, further gaining an insight into the operation of the system.
     \item \textbf{Unauthorised modification or theft of information}: 
      Bypassing controls an injecting modified information.
     \item \textbf{Denial of service or prevention of authorised access}:
      By injecting packets that cause the device to trigger a \ac{DOS}. 
     \item \textbf{Denial/Claim of action that took/not took place}: 
      By preventing legitimate responses from arriving the adversary can force records into an incomplete, or fictitious state.
 \end{itemize}

To perform these attacks on an industrial site without causing any adverse effects on the system would require a lot of planning and resources. The exact physical device would need to be tested, and running the same program code as the live system, while also ensuring the devices are networked similarly. However, the attack may also be performed blindly without all the information, yet the exact results may be unknown. 

% =========================================================================== %
\section{\uppercase{Methodology}}
\label{sec:method}
% =========================================================================== %

% Detail some assumptions

When performing these experiments we assume that a highly skilled adversary has compromised the network and has positioned themselves so they can monitor and inject packets at any point of the network. This is consistent with previously reported intrusions of state actors within industrial environments, where it is viable for an adversary to compromise the network infrastructure, such as switches and routers, and have this level of provenance.
% Epxlain the high level method that's common for each of the experiemtns. 
Each experiment is configured so that there is a: client, server, and attacker. The attacker is positioned so that they are in between the client and server, this is ensured by introducing an artificial delay of 500ms to the responding device. This allows us to validate the methods used to inject the forged packets. In an industrial setting, the responding device may be in a remote location, using several communication mediums to respond to the operator, such as radio or public internet, providing an attacker with multiple opportunities. Once the injection has been completed, a recorded packet capture is parsed by the three \acp{IDS}, and their results manually analysed.
% highlight the innovations and knowledge contribution of this work.
The experiments will focus more on the \ac{IEC104} protocol over HTTP since our aim is to prove the possibility of applying this attack on protocols other than HTTP\cite{haagsma_deep_2015}, and within a localised network.

% =========================================================================== %
\section{\uppercase{Experiments}}
\label{sec:exp}
% =========================================================================== %

We performed four different \ac{MotS} attacks on the two protocols, HTTP and \ac{IEC104}. The HTTP experiments inject a false response then redirect the client. The \ac{IEC104} experiments consider the injection of a recorded response, and a completely forged response. 

% --------------------------------------------------------------------------- %
\subsection{HTTP}
% --------------------------------------------------------------------------- %

HTTP is a stateless protocol and is generally closed after each request/response is completed, resulting in the TCP handshake has to be performed for each request. With \ac{MotS}, it is possible to target HTTP connections based on IP addresses, as well as a specific user by triggering on cookie headers \cite{gallagher_how_2014} and other identifiable data sent within a HTTP connection. The network layout consists of three hosts (Client, Server, and Attacker) and a switch. A 500ms artificial delay has been added to the server to ensure messages sent by the attacker will reliably arrive ahead of the legitimate response. When the client initiates a HTTP GET request, the attacker generates a forged packet based IP/TCP fields of the request. Then, the attacker reverses the direction of the IP/TCP source and destination addresses/ports, while the remainder of the IP header is unchanged. A random IP ID is generated, and the forged packet's TCP sequence number is set to the request's TCP acknowledgement, and the forged packet's TCP acknowledgement to the request's TCP sequence number. Finally, the forged payload is injected and all checksums calculated before being transmitted to the client. The legitimate response from the server arrives at the client afterwards, and is consequently dropped by the client.

\subsubsection{Experiment 1: Inject Response Page} The forged response contains a simple HTML page with a heading. Instead of the legitimate page, the attacker could inject malicious JavaScript, as seen by \cite{nakibly_website-targeted_2016}. The result of this exchange is that the client renders the forged HTTP response in the web browser, while the legitimate response is dropped without the browser being notified. The connection needs to be terminated by sending the TCP flags \verb|FIN ACK|, which shuts down the connection. It is worth noting that if the attack uses \verb|PUSH ACK| TCP flags, which are the flags set under normal circumstances, the TCP connection will remain open and the two payloads are rendered as one by the browser. This happens even if the HTTP header \verb|Connection: Close| is sent. 

\subsubsection{Experiment 2: Inject Redirect} This time the forged responses contain a "HTTP 301 Moved Permanently", instead of "HTTP 200 OK". This status line requires the header field \verb|Location|, along with a URI for redirection. Once received, the browser will automatically open and render the specified URI. This can be used to redirect the user to an attacker-controlled site, which has obvious security implications. The outcome is that the client follows the 302 redirection URI, and drops the legitimate response. Additionally, due to the nature of HTTP, the server records the Get request into the server logs, and is not aware that the client did not receive the response. Consequently, the only way to detect this happening is to monitor the network layer. 

% --------------------------------------------------------------------------- %
\subsection{\acl{IEC104}}
% --------------------------------------------------------------------------- %
% Background IEC104
\ac{IEC104} \cite{international_electrotechnical_commission_en_2006} is a protocol for transporting \ac{IEC101} frames over TCP. \ac{IEC101} is a plaintext telecontrol protocol designed for serial links. In these experiments we will create forged \ac{IEC104} packets with \ac{IEC101} payloads. The structure of an \ac{IEC104} packet, comprises two segments, the \ac{ASDU} and \ac{APCI}. The \ac{ASDU} contains the data to be exchanged, represented as Information Elements, and data about the number of information elements are contained within the \ac{DUI}. The \ac{APCI} is sent with all packets and is used to specify the start/end bits along with the \ac{ASDU} length. The protocol manages protection against loss and duplication of messages using a counter mechanism. Section 5.1 of EN 60870-5-104:2006 \cite{international_electrotechnical_commission_en_2006} specifies that a send sequence number N(S) and a receive sequence number N(R) are used based on the method defined in ITU-T X.25. Both sequence numbers are incremented by one for each \ac{APDU} and each direction. These are acknowledged by the receiver, by returning the receive sequence number N(R) to the sender using the supervisory format. Once acknowledged, the sender can remove the correctly transmitted \acp{APDU} from its buffer.

% Testbed setup 
\begin{figure}
	\centering
	\resizebox{.8\linewidth}{!}{\input{figures/mots-104-break-down.tex}}
	\caption{A generalised network diagram showing two enclaves: \ac{SCADA} and Process Control.}
	\label{fig:mots-104-break-down}
\end{figure}
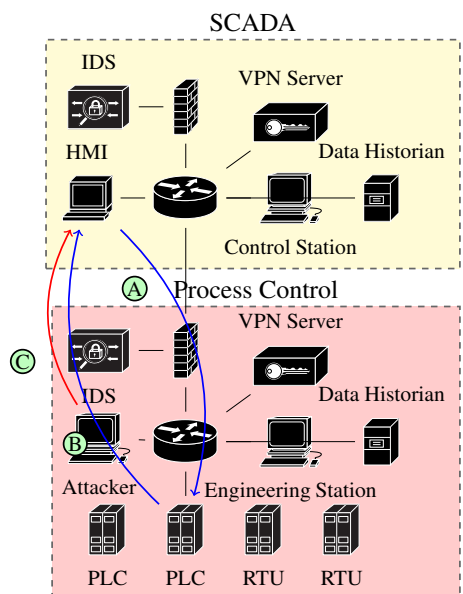

Figure~\ref{fig:mots-104-break-down} describes the \ac{SCADA} testbed used to perform the experiments. The attacker is located on the same Process Control enclave as the \ac{PLC}, while the victim, the \ac{HMI} is located within the \ac{SCADA} enclave. The attacker uses the port mirroring feature of the Process Control switch to monitor the network. The attacker builds the forged packet from the initial \ac{GI} request sent by the \ac{HMI} to the \ac{PLC} (Figure~\ref{fig:mots-104-break-down}: Step A), by switching the MAC and IP source/destination addresses, and switching the TCP sequence and acknowledgement numbers. Additionally, the send and receive counters for \ac{IEC104} are calculated to match the expected response. The attacker then waits for the \ac{PLC} to send an \verb|ActCon| (Figure~\ref{fig:mots-104-break-down}; Step B), before sending the forged response to the \ac{HMI} (Figure~\ref{fig:mots-104-break-down}; Step C). In the case of \ac{IEC104}, unlike the previously described HTTP attacks, the TCP flags remain as \verb|PUSH ACK|, which will be explained later.

% Attack sequence
These experiments are based on a \ac{GI} (GI and C\_IC\_NA\_1 will be used interchangeably) of a \ac{PLC} initiated from the \ac{HMI}. In normal operation, the \ac{PLC} would return its current status in response to a GI. The testbed is configured to generate and return data that simulates a real world response. This contains single-point, double-points, and step-position information. The objective of these experiments is to investigate how an attacker can forge a fake response to change the data viewed by a \ac{HMI} operator. 

\begin{figure}
  \centering
  \resizebox{.7\linewidth}{!}{\input{figures/104-con.tex}}
  \caption{Sequence diagram of a \acf{HMI} performing an \acf{GI} command on a \acf{PLC}. The dotted line shows the injected response.}
  \label{fig:iec104-seqence}
\end{figure}
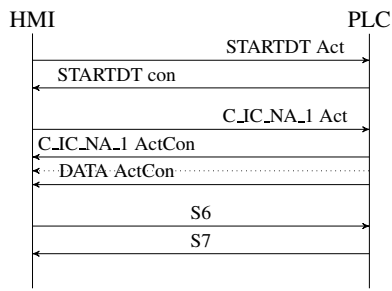

Figure~\ref{fig:iec104-seqence} is a sequence diagram detailing the steps of a \ac{GI}. After the TCP handshake is successful, the sender will initialise the connection for data transmission using the \verb|STARTDT = act| command, if successful the receiving station will respond with \verb|STARTDT = con|. Next, the sender performs a GI request, using an \ac{ASDU} frame (\verb|C_IC_NA_1 = act|), which is acknowledged by receiving end as \verb|C_IC_NA_1 = act con|, followed by the interrogation data. The dotted line is where the forged packets are injected within the sequence. Finally, the data is confirmed using the Supervisory format N(S), containing the number of frames received. 

\subsubsection{Experiment 3: Inject a Replayed Response}

This experiment will replay a pre-captured response from the \ac{PLC}, matching the same number of APCI/ASDU responses expected from the \ac{PLC}, so that an attacker can hide the physical state from the operator. Effectively, this attack will force the data visible to the operator to become out of sync with the physical process, in a similar way that Stuxnet hid real-time information from operators. Since this is a replay attack, the injected response will contain the same number of information elements, except the values will be reused from an earlier transmission. Listing~\ref{list:exp3-104} shows a packet capture of this experiment. The first packets are the TCP handshake (1-3), followed by the \verb|STARTDT act| and \verb|STARTDT con| messages (4-6) to start data transmission on this connection from the \ac{HMI} to the \ac{PLC}. The \ac{HMI} sends a GI (7), and at this point, the attacker generates the forged response. The \ac{PLC} then acknowledges the GI (8-9), triggering the attacker to send their forged response (10). The forged response is accepted by the \ac{HMI} (11), after which the legitimate GI response arrives (12), and is marked in the packet capture as \verb|TCP Spurious Retransmission|. The \ac{HMI} responds to the \ac{PLC} with a TCP ACK (13), which is marked as \verb|TCP Dup ACK|, since it was already used by the forged response. 

After the replayed packet has been accepted, the \ac{PLC} software hangs, then drops because the TCP sequence number becomes out of sync due to the injected segment. Although, as was the case in the HTTP experiments, it is possible to close the connection, the operation of \ac{IEC104} normally requires long lived connections. Deliberately bringing down this connection with a TCP reset could alert a system administrator to an issue or intrusion within the network. While the result is that the \ac{HMI} operator is now viewing stale data, the connection will be dropped and a redundant connection used in its place. The \ac{IEC104} standard states that several connections may be maintained with the remote device, and periodically refreshed using the \verb|STARTDT| command. Allowing the connection to degrade using the time out mechanism, overusing the TCP RST method, reduces the likelihood of triggering an alert.

% \noindent 
% \begin{minipage}{\linewidth}
\begin{lstlisting}[numbers=left,basicstyle=\tiny,caption={Packet Capture of IEC104 Experiment 3.},label=list:exp3-104]
10.50.50.103 > 10.50.50.99  TCP 74 39394 > 2404 [SYN] Seq=0 Win=29200 Len=0
 10.50.50.99 > 10.50.50.103 TCP 74 2404 > 39394 [SYN, ACK] Seq=0 Ack=1 Win=8192 Len=0
10.50.50.103 > 10.50.50.99  TCP 66 39394 > 2404 [ACK] Seq=1 Ack=1 Win=29312 Len=0
10.50.50.103 > 10.50.50.99  104apci 72 <- U (STARTDT act) 
 10.50.50.99 > 10.50.50.103 104apci 72 -> U (STARTDT con) 
10.50.50.103 > 10.50.50.99  TCP 66 39394 > 2404 [ACK] Seq=7 Ack=7 Win=29312 Len=0
10.50.50.103 > 10.50.50.99  104asdu 82 <- I (0,0) ASDU=1 C_IC_NA_1 Act     IOA=0
 10.50.50.99 > 10.50.50.103 104asdu 82 -> I (0,1) ASDU=1 C_IC_NA_1 ActCon  IOA=0
10.50.50.103 > 10.50.50.99  TCP 66 39394 > 2404 [ACK] Seq=23 Ack=23 Win=29312 Len=0
 10.50.50.99 > 10.50.50.103 104asdu 692 -> I (1,1) ASDU=1 C_IC_NA_1 ActCon  IOA=0 | -> I (2,1) ASDU=1 M_SP_NA_1 Inrogen IOA[60]=65583,... | -> I (3,1) ASDU=1 M_SP_NA_1 Inrogen IOA[44]=1120366,... | -> I (4,1) ASDU=1 M_DP_NA_1 Inrogen IOA[30]=561192,... | -> I (5,1) ASDU=1 M_ST_NA_1 Inrogen IOA[2]=1874019,... | -> I (6,1) ASDU=1 C_IC_NA_1 ActTerm IOA=0
10.50.50.103 > 10.50.50.99  TCP 66 39394 > 2404 [ACK] Seq=23 Ack=649 Win=30464 Len=0
 10.50.50.99 > 10.50.50.103 104asdu 692 [TCP Spurious Retransmission]  | -> I (1,1) ASDU=1 C_IC_NA_1 ActCon  IOA=0 | -> I (2,1) ASDU=1 M_SP_NA_1 Inrogen IOA[60]=65583,... | -> I (3,1) ASDU=1 M_SP_NA_1 Inrogen IOA[44]=1120366,... | -> I (4,1) ASDU=1 M_DP_NA_1 Inrogen IOA[30]=561192,... | -> I (5,1) ASDU=1 M_ST_NA_1 Inrogen IOA[2]=1874019,... | -> I (6,1) ASDU=1 C_IC_NA_1 ActTerm IOA=0
10.50.50.103 > 10.50.50.99  TCP 78 [TCP Dup ACK 11#1] 39394 > 2404 [ACK] Seq=23 Ack=649 Win=30464 Len=0
\end{lstlisting} 
% \end{minipage}

\subsubsection{Experiment 4: Inject Two Step Position Responses}

This experiment will investigate injecting two step position responses into the connection, rendering the \ac{HMI} with an incomplete view of the system. In Experiment 3, an old response was returned containing the full amount of expected data, however in Experiment 4 a forged response containing only two step position values, i.e. fewer than expected, will be returned. The motivation for this experiment is to determine what would happen when a smaller than expected response arrives on an active TCP connection. Note that in the HTTP experiments two payloads would be rendered as one if the connection was not terminated. Listing~\ref{list:exp2_104} shows a packet capture of this experiment. The first packets are the TCP handshake and IEC104 start data transfer acknowledgement (1-6). Next, the \ac{GI} request and acknowledgement are seen (7-9), followed by the forged packet (10-11). Note the payload is 120bits compared to the legitimate payload of 692bits (12). As with the previous experiment, the \ac{HMI} confirms the received frames using the Supervisory format (14), this time with the receive sequence number of 4, since there were 4 frames sent: Opening interrogation command ``ActCon''; two step positions; and Closing interrogation command ``ActTerm''. Subsequently, the connection keeps using the wrong TCP sequence number which eventually causes it to time out. 

% \noindent
% \begin{minipage}{\linewidth}
\begin{lstlisting}[numbers=left,basicstyle=\tiny,caption={Packet Capture of IEC104 Experiment 4.},label=list:exp2_104]
10.50.50.103 > 10.50.50.99  TCP 74 39440 > 2404 [SYN] Seq=0 Win=29200 Len=0
 10.50.50.99 > 10.50.50.103 TCP 74 2404 > 39440 [SYN, ACK] Seq=0 Ack=1 Win=8192 Len=0
10.50.50.103 > 10.50.50.99  TCP 66 39440 > 2404 [ACK] Seq=1 Ack=1 Win=29312 Len=0
10.50.50.103 > 10.50.50.99  104apci 72 <- U (STARTDT act) 
 10.50.50.99 > 10.50.50.103 104apci 72 -> U (STARTDT con) 
10.50.50.103 > 10.50.50.99  TCP 66 39440 > 2404 [ACK] Seq=7 Ack=7 Win=29312 Len=0
10.50.50.103 > 10.50.50.99  104asdu 82 <- I (0,0) ASDU=1 C_IC_NA_1 Act     IOA=0
 10.50.50.99 > 10.50.50.103 104asdu 82 -> I (0,1) ASDU=1 C_IC_NA_1 ActCon  IOA=0
10.50.50.103 > 10.50.50.99  TCP 66 39440 > 2404 [ACK] Seq=23 Ack=23 Win=29312 Len=0
 10.50.50.99 > 10.50.50.103 104asdu 120 -> I (1,1) ASDU=1 C_IC_NA_1 ActCon  IOA=0 | -> I (2,1) ASDU=1 M_ST_NA_1 Inrogen IOA[2]=1874019,... | -> I (3,1) ASDU=1 C_IC_NA_1 ActTerm IOA=0
10.50.50.103 > 10.50.50.99  TCP 66 39440 > 2404 [ACK] Seq=23 Ack=77 Win=29312 Len=0
 10.50.50.99 > 10.50.50.103 TCP 692 [TCP Out-Of-Order] 2404 > 39440 [PSH, ACK] Seq=23 Ack=23 Win=66560 Len=6
10.50.50.103 > 10.50.50.99  TCP 78 [TCP Dup ACK 11#1] 39440 > 2404 [ACK] Seq=23 Ack=77 Win=29312 Len=0
10.50.50.103 > 10.50.50.99  104apci 72 <- S (4) 
\end{lstlisting} 
% \end{minipage}

The PLC and the web server both had an artificial delay of 500ms to allow for the injected packets to reach the victim before the legitimate response. This is due to the limited scale of the laboratory testbed, and consequent low latency, this enabled the experiments to be developed without concern for the attacker to beat the legitimate packets. Nonetheless, in this testbed, when the artificial delays were completely removed, it was found that the forged response would arrive ahead of the legitimate response approximately 1 in 7 times. 

% --------------------------------------------------------------------------- %
\section{\uppercase{Detection}}
\label{sec:detection}
% --------------------------------------------------------------------------- %

Each of the four experiments were analysed offline using three of the most recent state of the art network \aclp{IDS}. These are Zeek 2.6.2 (formerly Bro), Snort 2.9.13-1, and Suricata 4.1.4. All of the engines were updated to the latest public rule sets as of May 2019, and were configured to parse offline packet capture files. Table~\ref{tab:nids_detection} shows details of the detection results for all three engines. Successful detection is recorded as either 'Yes', 'No', or 'Partial'. 'Yes' indicates an exact match on the correct packet, with the triggered rule returning an accurate description of the respective attack. 'Partial' means, for example, a rule triggered at the correct time but the reason for the alert, or the description provided, was not accurate. 'No' indicates no IDS alert was triggered. % IDSs and their MotS detection
\begin{table}
  \centering
  \caption{Network IDS detection results. (\CIRCLE) Yes (\Circle) No (\LEFTcircle) Partial }
  \label{tab:nids_detection}
  \begin{tabular}{lcccc}
    \toprule
     & \multicolumn{2}{c}{HTTP} & \multicolumn{2}{c}{IEC104}  \tabularnewline
    \cmidrule{2-3} \cmidrule{4-5}
    Experiment &  1 & 2 & 3 & 4 \tabularnewline
    \midrule
    Zeek     &\LEFTcircle &\LEFTcircle &\LEFTcircle  &\LEFTcircle \tabularnewline
    Snort    &\CIRCLE     &\LEFTcircle &\Circle      &\Circle     \tabularnewline
    Suricata &\CIRCLE     &\CIRCLE     &\Circle      &\CIRCLE     \tabularnewline
    \bottomrule
  \end{tabular}
\end{table}
% Zeek
Zeek could be considered the least accurate of the three engines in terms of its alerts. For the HTTP experiments, Zeek alerted with `\textit{FIN\_advanced\_last\_seq}'. This alert is triggered because of the server accepting the forged packet, which brings down the connection. For the IEC104 experiments Zeek alerts with `\textit{window\_recision}' on the exact injected packet. However the IDS was not triggering due to the specific MotS activity, rather it alerted because the TCP recv-window shrank by more than the amount of data being ACKed, as detailed in RFC793 Section 3.7, which is strongly discouraged. Therefore, although Zeek alerted in all four cases, the reason for each alert would not be obvious to an analyst monitoring the alerts. Zeek has built in support for detecting \ac{MotS} attacks since 2015, however, it is disabled by default and does not detect the experiments.
% Snort Detection
Snort alerted on the correct injected packet in the first HTTP experiment, with the message `\textit{INVALID CONTENT-LENGTH OR CHUNK SIZE}'' using the 'http\_inspect' module. The second HTTP experiment was also correctly detected, however, the alerts are due to the legitimate segments arriving after the connection is closed with `\textit{Data sent on stream not accepting data}' and `\textit{Reset outside window}'. These are seen in both experiments 1 and 2. The only event which Snort alerts for regarding the IEC104 injections are `\textit{Consecutive TCP small segments exceeding threshold}', which is due to the server attempting to renegotiate the connection, and causing a time out. Therefore this was considered as not specifically detecting the MotS attack.
% Suricata
Suricata had the most accurate detection of the three IDSs. With the HTTP experiments, the \verb|ACK| packets with the wrong sequence numbers are detected and, as with Snort, the first HTTP injection is detected with `\textit{SURICATA HTTP unable to match response to request}'. The most interesting alert is `\textit{SURICATA STREAM reassembly overlap with different data}', which exactly describes the MotS attack. This is displayed for all experiments, except experiment 3, which related to injecting a replayed response. Experiment 3 did not trigger any alerts from Suricata, because the forged payload was the same as the legitimate one.
% ===========================================================================%
\section{\uppercase{Conclusion}}
\label{sec:conclude}
% =========================================================================== 
% Detection
Further work should be performed for each of the NIDS to include detection alerts for the \ac{MotS} class of attacks. While Suricata can detect this, and notify the operator with a somewhat descriptive name, it fails to articulate the issue as a malicious action. Network \ac{IDS} for deployment within \ac{OT} networks must have support for the industrial protocols, otherwise, detection of these kinds of attacks and others will go unnoticed. 
% Mitiagtions
As discussed in the introduction, zero-trust networks, \acp{VPN} (to an extent), and TLS are the best mitigations of this class of attack. \ac{IEC104} is has a companion standard, IEC 62351, detailing securing end to end communication, however in internal networks, these defences are often not deployed. In part due to a lax security mindset, or the additional complexity and risk of deployment. While this class of attack is mitigated for much of the public internet via TLS, critical network operators need to be aware of the potential damage this attack may have on a system. Although deploying a mitigation approach such as zero-trust networking is the gold standard, many network operators are not in a position to do so. Threat actors, such as nation states have used this method successfully over several years, and it seems likely this approach will continue to be exploited in the future. Especially in the ICS domain, where unauthenticated local traffic such as IEC104 is still commonplace. As the experiments with Zeek, Snort and Suricata have shown, further work is required to provide these network IDS platforms with the detection rules and mechanisms capable of accurately detecting MotS being exploited by an intruder.
% \section*{A Note on reproducibility}
% The experiments and results from this paper can be found online\footnote{\url{https://github.com/PMaynard/mots}.}, which contain all instructions and code necessary to replicate these experiments. 
\bibliographystyle{apalike}
{\small
\bibliography{MOTS}}
\end{document}

%% file: 00acronyms.tex
\acro{5-Tuple}{Protocol, IP src/dst and TCP Port src/dst}
\acro{ABB}{ASEA Brown Boveri}
\acro{AGC}{Automatic Generation Control}
\acro{ANN}{Artificial Neural Network}
\acro{APCI}{Application Protocol Control Information}
\acro{APDU}{Application Protocol Data Unit}
\acro{APH}{Acyclic Phase-Type}
\acro{API}{Application Program Interface}
\acro{ARP}{Address Resolution Protocol}
\acro{ASDU}{Application Service Data Unit}
\acro{ATTCK}{Adversarial Tactics, Techniques \& Common Knowledge} % TODO: Short version should have an apersand. - ATT&CK
\acro{AWS}{Amazon Web Services}
\acro{BACNet}{Building Automation and Control}
\acro{BAS}{Building Automation System}
\acro{BDMP}{Boolean Logic Driven Markov Processes}
\acro{BGP}{Border Gateway Protocol}
\acro{BPF}{Berkeley Packet Filter}
\acro{BSI}{Bundesamt für Sicherheit in der Informationstechnik}
\acro{BTC}{Baku-Tbilisi-Ceyhan}
\acro{C2}{Command and Control}
\acro{CAR}{Cyber Analytics Repository}
\acro{CERT}{Computer Emergency Response Team}
\acro{CIA}{Confidentiality Integrity and Availability}
\acro{CIP}{Common Industrial Protocol}
\acro{CIP}{Critical Infrastructure Protection}
\acro{CLI}{Command Line Interface}
\acro{CNI}{Critical National Infrastructure}
\acro{CoPNet}{Colored Petri Net}
\acro{COT}{Cause of Transmission}
\acro{CPNI}{Centre for the Protection of National Infrastructure}
\acro{CPS}{Cyber Physical System}
\acro{CPS}{Cyber Physical System}
\acro{CPT}{Conditional Probability Tables}
\acro{CVE}{Common Vulnerabilities and Exposures}
\acro{CVSS}{Common Vulnerability Scoring System}
\acro{CybOX}{Cyber Observable eXpression}
\acro{CySeMoL}{Cyber Security Modelling Language}
\acro{DCS}{Distributed Control System}
\acro{DHCP}{Dynamic Host Configuration Protocol}
\acro{DMZ}{Demilitarised Zone}
\acro{DNP3}{Distributed Network Protocol 3}
\acro{DNS}{Domain Name Service}
\acro{DOS}{Denial of Service}
\acro{DQU2}{Duqu 2.0}
\acro{DUI}{Data Unit Identifier}
\acro{ELK}{Elastic Search, Logstash and Kibanna}
\acro{EM}{Expect Maximisation}
\acro{ENIP}{EtherNet/IP}
\acro{GCHQ}{Government Communications Headquarters}
\acro{GMM}{Gaussian Mixture Model}
\acro{GOOSE}{Generic Object Oriented Substation Events}
\acro{GSM}{Global System for Mobile Communications}
\acro{HMI}{Human Machine Interface}
\acro{HTTP}{Hypertext Transfer Protocol} % Included for completness
\acro{HVAC}{Heating, Ventilation, and Air Conditioning}
\acro{ICMP}{Internet Control Message Protocol} % Included for completness
\acro{ICN}{Information Centric Networking}
\acro{ICS-CERT}{ICS Computer Emergency Response Team}
\acro{ICS}{Industrial Control System}
\acro{ICS-KC}{ICS Cyber Kill Chain}
\acro{IDMEF}{Intrusion Detection Message Exchange Format}
\acro{IDS}{Intrusion Detection System}
\acro{IEC101}{IEC 60870-5-101}
\acro{IEC104}{IEC 60870-5-104}
\acro{IEC61508}{IEC61508}
\acro{IEC61850}{IEC61850} 
\acro{IEC}{International Electrotechnical Commission}
\acro{IED}{Intelligent Electronic Device}
\acro{IOA}{Information Object Address} 
\acro{IOC}{Indicator of Compromise} \acrodefplural{IOC}[IOCs]{Indicators of Compromise}
\acro{IoT}{Internet of Things}
\acro{IOT}{Internet of Things}
\acro{IP}{Internet Protocol} % Included for completness
\acro{IT}{Information Technology} % Included for completness
\acro{IPS}{Intrusion Prevention System}
\acro{IR}{Incident Response}
\acro{JSON}{JavaScript Object Notation}
\acro{KM}{K-Means}
\acro{KNN}{k-Nearest Neighbours}
\acro{LAN}{Local Area Network}
\acro{LDS}{Leak Detection System}
\acro{LSTM}{Long-Short Term Memory}
\acro{MAC}{Media Access Control}
\acro{MBR}{Master Boot Record}
\acro{MCC}{Matthews Correlation Coefficient}
\acro{MitM}{Man-in-the-Middle}
\acro{MMS}{Manufacturing Message Specification}
\acro{Modbus}{Modbus-TCP}
\acro{MPTCP}{Multipath TCP}
\acro{MTTC}{Mean-Time-to-Compromise}
\acro{NB}{Naive Bayesian}
\acro{NCSC}{National Cyber Security Centre}
\acro{NDA}{Non-Disclosure Act} 
\acro{NERC}{North American Electric Reliability Corporation}
\acro{NIST}{National Institute of Standards and Technology}
\acro{NSM}{Network Security Monitoring}
\acro{OCSVM}{One-Class SVM}
\acro{OPC}{OLE for Process Control}
\acro{OPC-UA}{OPC Unified Architecture}
\acro{OSGP}{Open Smart Grid Protocol}
\acro{OSI}{Open Systems Interconnection}
\acro{OT}{Operational Technology}
\acro{PCAP}{Packet Capture}
\acro{PCA-SVD}{Component Analysis using Singular Value Decomposition}
\acro{PCN}{Process Control Network}
\acro{PLC}{Programmable Logic Controller}
\acro{PRECYSE}{Prevention, protection, and reaction to cyber attacks to critical infrastructures}
\acro{PROFINET}{Process Field Net}
\acro{PSTN}{Public Switched Telephone Network}
\acro{RAT}{Remote Access Trojan}
\acro{RBF}{Radial Basis Function}
\acro{RTU}{Remote Terminal Unit}
\acro{S7}{S7 Communication}
\acro{SAND}{Attack Trees with Sequential Conjunction}
\acro{SCADA}{Supervisory Control And Data Acquisition}
\acro{SDN}{Software Defined Networking}
\acro{SIL}{Safety Integrity Level}
\acro{SIQ}{Single-point information with quality descriptor}
\acro{SIS}{Safety Instrumented System}
\acro{SMB}{Server Message Block}
\acro{SPI}{Status ON}
\acro{STIX}{Structured Threat Information Expression}
\acro{STP}{Spanning Tree Protocol}
\acro{SVM}{Support Vector Machine}
\acro{SysML}{Systems Modelling Language}
\acro{SysML}{Systems Modelling Language}
\acro{TCP}{Transmission Control Protocol} % Included for completness
\acro{TLS}{Transport Level Security}
\acro{UDP}{User Datagram Protocol} % Included for completness
\acro{UPS}{Uninterruptible Power Supply} \acrodefplural{UPS}[UPSs]{Uninterruptible Power Supplies}
\acro{VCG}{Vulnerability Cause Graph}
\acro{VLAN}{Virtual LAN} % Included for completness
\acro{VPN}{Virtual Private Network}
\acro{WEKA}{Waikato Environment for Knowledge Analysis}
\acro{WPAD}{Web Proxy Auto-Discovery}
\acro{XSS}{Cross Site Scripting}
\acro{MotS}{Man-on-the-Side}
\acro{ACL}{Access Control List}
\acro{TTL}{Time To Live}
\acro{GI}{General Interrogation}
\acro{ActCon}{???Accept Connection???}
\acro{PoC}{Proof of Concept}

%% file: figures/mots-sequence.tex
% arara: pdflatex 

% \documentclass[tikz]{standalone}
% \usetikzlibrary{calc,positioning,arrows}
% \begin{document}

\begin{tikzpicture}[node distance=1.5cm,auto,>=stealth']
    % set up nodes
    \node[] (client) {\large{Client}};
    \node[right = of client] (attacker) {\large{Attacker}};
    \node[right = of attacker] (server) {\large{Server}};
    \node[below of=client, node distance=3.5cm] (client_ground) {};
    \node[below of=attacker, node distance=3.5cm] (attacker_ground) {};
    \node[below of=server, node distance=3.5cm] (server_ground) {};
    % Horizontal Lines
    \draw (client) -- (client_ground);
    \draw (attacker) -- (attacker_ground);
    \draw (server) -- (server_ground);
    % Events
    \draw[->] ($(client)!0.25!(client_ground)$) -- node[above,scale=1, near start]{1. Initiate Connection} ($(server)!0.25!(server_ground)$);
    \draw[<-] ($(client)!0.55!(client_ground)$) -- node[above,scale=1, near end]{2. Send forged response (\checkmark)} ($(attacker)!0.55!(attacker_ground)$);
    \draw[<-] ($(client)!0.80!(client_ground)$) -- node[above,scale=1, midway]{3. Send legitimate response ($\times$)} ($(server)!0.80!(server_ground)$);
\end{tikzpicture}

% \end{document}

% Thanks to: 
% https://tex.stackexchange.com/questions/207240/drawing-simple-sequence-diagram

%% file: figures/highlevel-ics-overview.tex
% \documentclass{standalone}
% \usepackage{tikz}
% \usetikzlibrary{positioning}
% \usetikzlibrary{fit}
% \begin{document}

\begin{tikzpicture}[
	every node/.style={
		rectangle,
		% draw=black!60,
		thick,
		text width=2.3cm,
		align=center},
	node distance=5mm
]

\node (internet) at (1.5,1.3) [ellipse]{DMZ};

\node (b1) at (0,0) {Enterprise Workstations};
\node (b2) [right=of b1] {Reporting};

\begin{scope}[on background layer]
\node[fill=green!20,draw=black,thin,dashed,fit=(b1) (b2),label={[align=left, text width=2.3cm]above left:Business}] (b) {};
\end{scope}
\node(space) [fill=none,draw=none,below=of b1] {};

\node (s1) [below=2pt of space] {HMI};
\node (s2) [right=of s1] {Data Historian};

\begin{scope}[on background layer]
\node[fill=yellow!20,draw=black,thin,dashed,fit=(s1) (s2),label={[align=left, text width=2.5cm]above left:SCADA}] (s) {};
\end{scope}
\node(space2) [fill=none,draw=none,below=of s1] {};

\node (p1) [below=2pt of space2] {PLC};
\node (p2) [right=of p1] {RTU};

\begin{scope}[on background layer]
\node[fill=red!20,draw=black,thin,dashed,fit=(p1) (p2),label={[align=left, text width=2.5cm]above left:Process Control}] (p) {};
\end{scope}
\node (pd) [below=of p] {Physical Domain};

% Lines 
\draw[-] (b) -- (internet);
\draw[-] (b) -- (s);
\draw[-] (s) -- (p);
\draw[-] (p) -- (pd);
\draw[-] (p) -- (pd);

\end{tikzpicture}

% \end{document}  

%% file: figures/mots-104-break-down.tex
% arara: pdflatex 

% \documentclass{standalone}
% \usepackage{tikz}
% \usepackage{epstopdf}

% \usetikzlibrary{positioning}
% \usetikzlibrary{fit}
% \usetikzlibrary{graphs}
% \usetikzlibrary{backgrounds}

% \begin{document}
% \graphicspath{ {./cisco/} }

\begin{tikzpicture}[
	node distance=4mm,
	boundry/.style={
		draw=black!60, 	thick
	}
]

%%%%%%%%%%%%%%%%%%%%%%%%%%%%%%%%%%%%%%%%%%%%%%%%%%%%%
% SCADA
\node (scadafirewall)  											{\includegraphics[scale=.8]{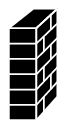}};
\node (scadarouter) 	[below=of scadafirewall]				{\includegraphics[scale=.8]{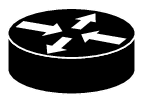}};
\node (scadaids) 		[left=of scadafirewall,label={[name=idsL] above:IDS}]	{\includegraphics[scale=.8]{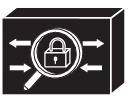}};

\node (scadavpn) [above right=of scadarouter,label=above:VPN Server] 	{\includegraphics[scale=.8]{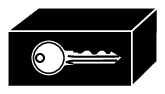}};
\node (hmi) 	[left=of scadarouter,label={[name=hmiL] above:HMI}] 		{\includegraphics[scale=.8]{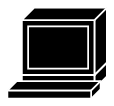}};
\node (scadaengineering) 	[right=of scadarouter,label={[name=scadaengineeringL] below:Control Station}] 		{\includegraphics[scale=.8]{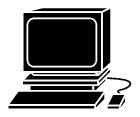}};
\node (scadadatahist) 	[right=of scadaengineering,label={[name=scadadatahistL] above:Data Historian}] 		{\includegraphics[scale=.8]{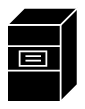}};

\begin{scope}[on background layer]
\node[boundry, fill=yellow!20, dashed, fit=(scadafirewall) (scadarouter) (scadaids) (idsL) (scadavpn) (hmi) (hmiL) (scadaengineeringL) (scadadatahist) (scadadatahistL), label={[font=\large]above:SCADA}] (scada) {}; 
\end{scope}
%%%%%%%%%%%%%%%%%%%%%%%%%%%%%%%%%%%%%%%%%%%%%%%%%%%%%
% Process Control
\node (pcfirewall)  [below=40pt of scadarouter] 				{\includegraphics[scale=.8]{cisco/firewall}};
\node (pcrouter) 	[below=of pcfirewall]						{\includegraphics[scale=.8]{cisco/router}};
\node (pcids) 		[left=of pcfirewall,label=below:IDS]		{\includegraphics[scale=.8]{cisco/ACS}};
\node (pcattacker) 	[below=of pcids,label=below:Attacker]		{\includegraphics[scale=.8]{cisco/workstation}};

\node (pcvpn) [above right=of pcrouter,label=above:VPN Server] 								{\includegraphics[scale=.8]{cisco/vpn_gateway}};
\node (engineering) 	[right=of pcrouter,label=below:Engineering Station] 				{\includegraphics[scale=.8]{cisco/workstation}};
\node (datahist) 	[right=of engineering,label={[name=datahistL] above:Data Historian}] 	{\includegraphics[scale=.8]{cisco/file_server}};

\node (plc) 	[below=of pcrouter,label={[name=plcL] below:PLC}] 	{\includegraphics[scale=.8]{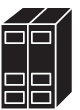}};
\node (plc1) 	[left=of plc,label={[name=plc1L] below:PLC}] 		{\includegraphics[scale=.8]{cisco/standard_host}};

\node (rtu) 	[right=of plc,label={[name=rtuL] below:RTU}] 		{\includegraphics[scale=.8]{cisco/standard_host}};
\node (rtu1) 	[right=of rtu,label={[name=rtu1L] below:RTU}] 		{\includegraphics[scale=.8]{cisco/standard_host}};

\begin{scope}[on background layer]
\node[boundry, fill=red!20, dashed, fit=(pcfirewall) (pcrouter) (pcids) (pcattacker) (engineering) (datahist) (datahistL) (plc) (plcL) (rtu) (rtuL) (plc1) (plc1L) (rtu1) (rtu1L), label={[font=\large]above:Process Control}] (pc) {}; 
\end{scope}

%%%%%%%%%%%%%%%%%%%%%%%%%%%%%%%%%%%%%%%%%%%%%%%%%%%%%
% Paths

% SCADA
\graph [use existing nodes] {
	scadaids -- scadafirewall ;
	scadarouter -- {scadafirewall, scadavpn, hmi, scadaengineering, scadadatahist}; 
};
% Proccess
\graph [use existing nodes] {
	pcids -- pcfirewall -- scadarouter;
	pcrouter -- {pcfirewall, pcrouter, pcvpn, engineering, datahist, plc, pcattacker};
};

% Attacker lines
\graph [use existing nodes, 
edge quotes={left, near start, black, fill=green!25, inner sep=.5pt, outer sep=10pt, draw, circle}, 
edge={bend left, blue, thick}] {
	hmi -> ["A"] plc;
	plc -> ["B"] hmi;
	pcattacker -> ["C", red] hmi;
};

\end{tikzpicture}

% \end{document}  

%% file: figures/104-con.tex
% arara: pdflatex 

% \documentclass[tikz]{standalone}
% \usetikzlibrary{calc,positioning,arrows}
% \begin{document}

\begin{tikzpicture}[node distance=5cm,auto,>=stealth']
    % set up nodes
    \node[] (hmi) {\large{HMI}};
    \node[right = of hmi] (plc) {\large{PLC}};
    \node[below of=hmi, node distance=5cm] (hmi_ground) {};
    \node[below of=plc, node distance=5cm] (plc_ground) {};
    % Horizontal Lines
    \draw (hmi) -- (hmi_ground);
    \draw (plc) -- (plc_ground);
    % Events
    \draw[->] ($(hmi)!0.15!(hmi_ground)$) -- node[above,scale=1, near end]{STARTDT Act} ($(plc)!0.15!(plc_ground)$);
    \draw[<-] ($(hmi)!0.25!(hmi_ground)$) -- node[above,scale=1, near start]{STARTDT con} ($(plc)!0.25!(plc_ground)$);

    \draw[->] ($(hmi)!.4!(hmi_ground)$) -- node[above,scale=1, near end]{C\_IC\_NA\_1 Act} ($(plc)!.4!(plc_ground)$);
    \draw[<-] ($(hmi)!.5!(hmi_ground)$) -- node[above,scale=1, near start]{C\_IC\_NA\_1 ActCon} ($(plc)!.5!(plc_ground)$);

    \draw[<-,dotted] ($(hmi)!.55!(hmi_ground)$) -- node[above,scale=1, near start]{} ($(plc)!.55!(plc_ground)$);
    \draw[<-] ($(hmi)!.6!(hmi_ground)$) -- node[above,scale=1, near start]{DATA ActCon} ($(plc)!.6!(plc_ground)$);

    \draw[->] ($(hmi)!.75!(hmi_ground)$) -- node[above,scale=1, midway]{S6} ($(plc)!.75!(plc_ground)$);
    \draw[<-] ($(hmi)!.85!(hmi_ground)$) -- node[above,scale=1, midway]{S7} ($(plc)!.85!(plc_ground)$);
\end{tikzpicture}

% \end{document}

% Thanks to: 
% https://tex.stackexchange.com/questions/207240/drawing-simple-sequence-diagram